\documentclass[12pt]{article}

\usepackage{epsfig}
\usepackage{amsfonts}
\usepackage{amsopn}
\usepackage{amsmath}
\usepackage{verbatim,amsthm}

\title
{
\vskip-50 pt
\begin{flushright}
\normalsize\rm NORDITA-2018-072
\end{flushright}
\vskip 20 pt  
$p$-Branes with $AdS_{p+1}$ vacuum as models of $R^2$ gravity  
}
\author{
 A. A. Zheltukhin $^{a,b}$\thanks{e-mail: aaz@nordita.org
 } 
  \\ \\
$^a$ Kharkov Institute of Physics and Technology, 
Kharkov, 61108, Ukraine \\  
$^b$ Nordita, KTH Royal Institute of Technology and Stockholm University\\
SE 106 91 Stockholm, Sweden}

\begin{document}
\maketitle
\begin{abstract}
 
 Branes with {\it constant mean curvature} of their hyper-worldsheets 
 of codimension  1 are treated as the Nambu-Goldstone fields 
 of the broken Poincare symmetry. 
 Mapping of their action into quadratic curvature gravity action 
 with spontaneously generated gravity, is shown. 
Equation for the  brane  potential extremals and its solution describing 
 hyper-ws of {\it constant curvature} are found. 
 For membranes in $\mathbf{R}^{1,3}$  this extremum is shown to be 
  a {\it saddle} 3-dim. hypersurface which defines classically 
  instable vacuum. 

\end{abstract}

\section{Introduction}

 Invention of the Green-Schwarz superstring and kappa symmetry led to the construction of a renormalizable ten-dimensional theory of all fundamental interactions including quantum gravity \cite{GSW}.
  The local kappa symmetry acting on two-dimensional worldsheet of superstring is a key element of the theory which provides the balance between the bosonic and fermionic degrees of freedom. The very existence of this symmetry imposes essential restrictions on the dimension of the ambient supersymmetric space-time in which strings move. The admissible dimensions of the space-time are $D=3,4,6,10$, but only $D=10$ turns out to be consistent in the quantum picture resulting in five possible renormalizible theories \cite{GS}. The latter are known as type I, type IIA, type IIB  supplemented by two theories originating from heterotic strings invariant under $SO(32)$ and $E8 \times E8$ gauge symmetries, respectively. These five theories are based on  different types of strings:  type I theory includes both open and closed strings, while  types IIA and IIB include only closed strings. All these theories are entangled by $S$ and $T$ duality transformations.  In particular, $S$-duality transforms type $I$ theory to the $SO(32)$ heterotic one, and type $IIB$ theory into itself. The latter is transformed into type IIA theory under $T$-duality transformations \cite{PolWit}. These theories are believed to be limiting cases of one unknown theory which consistently describes quantum gravity. 
Herewith, the $D=10, N=2a$ supergravity, in particular, arises as  low energy limit 
of Type IIA superstring.  In this limit which  corresponds  to the infinitely large string tension the discussed five theories yield different 10-dimensional versions of supergravity. The later belongs to field theories associated with point-like particles treated as idealized  geometric objects with the dimension $p=0$. So, the transition from point-like particles to one-dimensional extended objects ($p=1$) cures $D=10$ supergravity from ultraviolet divergencies. However, this scenario works only for the dimension $D=10$ that implies the Kaluza-Klein compactification of six excess coordinates in order to restore the physical dimension $D=4$.
But such a compactification creates the Landscape problem \cite{Susk_L} 
that means loss of uniqueness in the choice of the physically relevant universes
 describing $SU(3)\times SU(2)\times U(1)$ invariant interactions and three families of quarks and leptons \cite{Lind}. 
 Unfortunately, the list of such universes contains billions of candidates. Another problem is  the existence of unique $N=1$ supergravity in eleven dimensions which cannot be derived from the superstring theories in 10-dimensional space-time. At the same time the 10-dimensional IIA superstring is derived from 11-dimensional supermembrane \cite{BST} through simultaneous dimensional reduction in both $D=11$ space-time and 3-dimensional  $(p=2)$ supermembrane worldsheet \cite{DHIS}. 
In view of this observation, the $N=2a,  D=10$ supergravity is also derived 
from the $N=1,  D=11$ supermembrane. As a result, supermembranes and superstrings get equal status in the frame of mysterious 11-dimensional $M$-theory \cite{Wit_M}. 
But the story does not end there, in view of the tendency to further growth 
of the dimension $p$ of relevant extended objects as shown in \cite{AETW}.
Therein, the set of the numbers $p$ and $D$,  ensuring the presence 
of the kappa symmetry on a $(p+1)$-dimensional hyper-worldsheet embedded 
into $D$-dimensional Minkowski space, was revealed. 
This set includes, in particular, the values $p=2,  D=4,5,7,11$; $p=3,  D=6,8$; $p=4,  D=9$ and $p=5,  D=10$ corresponding to the embeddings of membranes and  3-, 4-, 5-branes into the mentioned higher-dimensional Minkowski spaces. 
The pair $p=5, D=10$ describes the heterotic 5-brane associated with the 
soliton solution of the heterotic string equations found in \cite{Str}. 
 In $M$-theory the supermembrane and super 5-brane are called $M2$ and $M5$ branes. 
M-branes together with superstrings are treated as fundamental constituents of supersymmetric 11-dimensional $M$-theory unifying gravity with other fundamental forces. Thus, both branes and strings play key roles in theoretical understanding of fundamental interactions,  phenomenology of elementary particle physics, black holes, dark matter and AdS/CFT correspondence. A comprehensive review of these and other problems of contemporary physics in which branes play a vital role may be found in \cite{Duf}.
The breakthrough role of branes in attempts to reveal new physics sharpens the problem of their quantum consistency. Solution of this problem has to shed a new light on quantization of gravity.  Until now quantization of extended objects remains an open problem even for the case of (super)membranes (p=2) \cite{DIPSS}. This uncertainty in the quantum status of the brane paradigma requires new tools for  analyzing its viability.  A weak progress in this matter is explained by a complicated non-linear character of brane equations and constraints. The non-linearity problem is typical of the theories unifying the Yang-Mills, gravitational and other fields covariant under diffeomorphisms, gauge and internal symmetries. To solve the problem, we need to deepen understanding of the structure of brane non-linearities. 

Here we assume that the non-linearities can be converted into the geometric 
structures known from gravity and gauge theories.  For branes, the proof of 
this assumption allows to implement the BRST-BFV quantization and others methods 
used for general non-linear systems. We expect that the non-linearities of ordinary $p$-branes 
can be described by a combination of scalars built from the curvature tensor of $(p+1)$-dimensional hyper-worldsheets (h-ws) \footnote{To be short we will further use the abbreviation h-ws for hyper-worldsheet(s).}.
 It implies a map of non-linear terms into those known from $(p+1)$-dim. $f(R)$ gravity which generalizes the Starobinsky 4-dim. model \cite{Strb}.
 Construction of such a map will also give an additional motivation for the hypothesis that our world is a 3-brane embedded into higher-dimensional spaces \cite{RS}, \cite{RZh1}, \cite{AADD}, \cite{DGP}.

The inherent non-linearities prevent the use of string harmonic 
oscillators for brane quantization. On the other hand, it is known that 
for some non-linear systems the transition to new dynamical variables, similar to
the action-angle variables, permits to linearize equations of motion and to solve them.
Then, the general solution makes it possible to perform quantization in terms of the 
 initial data. Such a possibility necessitates search for  variables alternative 
to the world vectors of brane h-ws. 
In string theory Regge and Lund \cite{RL} used
 the first and the second fundamental forms of string worldsheets 
embedded into 4-dim. Minkowski space as new variables 
(see also \cite{Omn}, \cite{BN}). Generalization of this idea for strings 
in $D$-dim. Minkowski space, accompanied with treatment of these variables 
as gauge multiplets, allowed to transform string equations into an exactly
 integrable system of PDEs \nocite{Zgau2, Zgau2'} [21,22].
So, it is natural to use such geometric variables for p-branes to verify
 whether brane non-linearities are integrable. 
 For this purpose we apply the renowned differential-geometric approach describing
 hypersurfaces embedded into Riemannian spaces based on the Maurer-Cartan structure
 equations \nocite{Car, Eisn, JS, ShSYa} [23 - 26]. 
This approach was represented in terms of gauge theory and spontaneous symmetry
 breaking in \cite{Z_rmp} for the Dirac (fundamental) p-branes sweeping minimal h-ws.  
 The spontaneous symmetry breaking (see \cite{BB} and refs. there) 
appears due to embedding of hypersurfaces in space-time.  
In the case a h-ws $\Sigma_{p+1}^{min}$ swept by a fundamental $p$-brane \cite{Dirb}
 in $D$-dim. Minkowski space causes spontaneous  breaking of its Poincare
symmetry  $ISO(1, D-1)$ to the subgroup $ISO(1,p)\times SO(D-p-1)$ 
 \nocite{GKW, GM, GKP, BZnul} [30 - 33].
The breaking yields the Nambu-Goldstone (N-G) fields identified with the coefficients
 $l_{\mu\nu}^{a}$ of the second fundamental form of $\Sigma_{p+1}^{min}$ with arbitrary codimention. These coefficients form  a N-G tensor multiplet of the local
 group $SO(D-p-1)$. The Y-M multiplet $B_{\mu}^{ab}$ of this group, together with 
the N-G multiplet and the metric $g_{\mu\nu}$, compose a set of $(p+1)$-dim. h-ws fields. 
Their invariant action $S_{Dir}$ and the potential $V_{Dir}(l,g)$ 
built in \cite{Z_rmp} are  given by Eqs. (\ref{actnl}) and (\ref{solVl}) 
from Secton 2. The potential encodes the non-linearities of fundamental 
 p-branes in terms of $l_{\mu\nu}^{a}$ and $g_{\mu\nu}$.
 So, $V_{Dir}(l,g)$ expressed through the curvature tensor of $\Sigma_{p+1}^{min}$ 
has to uncover the required geometric structure of brane non-linearities.

This scheme was realized in \cite{Z_mpla} for fundamental $p$-branes 
sweeping minimal h-ws with codimension 1 (i.e. $D=p+2$).  In this case the gauge field 
$B_{\mu}^{ab}$ is absent. The Dirac membrane $p=2, D=4$ 
 is an example of such branes. It describes the bosonic sector of the 
supermembrane from the above discussed list of branes invariant under 
the kappa-symmetry \cite{AETW}. 
 The case $p=3, D=5$ selects fundamental $3$-branes living in 5-dim. Minkowski space.
 The $3$-brane  potential  follows from Eq. (\ref{solVl}) 
\begin{eqnarray}\label{Vred3}
V_{Dir}|_{p=3}= [- \frac{1}{2}Sp(l^{2})Sp(l^{2})+ {\rm c_{p}}]|_{p=3}. 
\end{eqnarray}
 As shown in \cite{Z_mpla},  $V_{Dir}$ is expressed through the squared scalar curvature $R$ of the h-ws $\Sigma_{p+1}^{min}$ by Eq. (\ref{Vred}) from Section 3 
\footnote{The formula (\ref{Vred3}) is the same for any  $(p+1)$-dim. minimal 
h-ws $\Sigma_{p+1}^{min}$ with codim 1.}. 
It means that action (\ref{actnl}) with codim 1 encodes a particular case 
of $(p+1)$-dim. gravity theories quadratic in curvature \nocite{Frad, Zwi, LPPS, Stel} [35 - 38]. 
So, the transition to new geometric variables uncovers the structure 
of non-linearities of fundamental branes with codim 1 and proves their similarity 
to non-linearities of $f(R)$  theories of gravity. 

For $c_{p}=0$ action $S_{Dir}$ (\ref{actnl}) contains only one coupling 
constant $\kappa_{p}\sim T_{p}^{\frac{3-p}{2(p+1)}}$, where $T_{p}$ is the tension of $p$-branes. The power law shows three different regimes of the behavior of $k_{p}$ as a function of the tension and, consequently, the energy corresponding to the cases $p<3$,  $p>3$ and $p=3$. These regimes testify to the  presence of the phases of asymptotic freedom and confinement and their links with inflation and collapse of the branes    \footnote{Exact solutions for p-brane collapse were earlier obtained
 in \nocite{TrZ, Zcol,Zcol'} [39-41].}. 
	
The phase with $p=3$ is peculiar because the coupling constant $k_{3}$ becomes 
dimensionless, so that $S_{Dir}|_{p=3}$ is scale-invariant. 
The latter forbids the Hilbert-Einstein (H-E) term because it breaks the scale symmetry.
	To hold the desired connection of $3$-branes with our $4$-dim. world,  we have to modify the potential of fundamental branes to activate the H-E term \cite{CLNY}. 
	
Here we find that the same problem interferes with 4-dim. scale-invariant and renormalizable theories of gravity quadratic in curvature (see \nocite{Dir, Adl, Zee,  DFF, Smol, Perv, JoNo} [43 - 49] and refs. there). 
The latter are converted to gravity theories containing the H-E term 
by implementing the idea of spontaneously generated gravity.
 This idea is illustrated by a simple example using the scale-invariant action \cite{Zee}
	\begin{eqnarray}\label{sptgrv}
	A= \int d^{4}x\sqrt{|g|}\, 
\{ \frac{\alpha}{2}\varphi^{2}R 
+   \frac{1}{2}\nabla_{\mu}\varphi\nabla_{\nu}\varphi g^{\mu\nu} - V(\varphi,g) \}.
	\end{eqnarray}  
Action (\ref{sptgrv}) contains  a scalar $\varphi$ and a dimensionless constant $\alpha$. 
The potential $V(\varphi,g)$ is assumed to have a deep minimum at $\varphi_{o}=v$ which 
provides vacuum expectation value (vev) $v$ for $\varphi$. 
Expansion of the first term in (\ref{sptgrv}) around the minimum generates 
 the H-E term with the Newton constant $G_{N}\approx\frac{1}{\alpha v^2}$.
 So, the scale symmetry of  (\ref{sptgrv}) is spontaneously broken that results 
in a 4-dim. gravity in the low energy limit. 
 On the contrary, in the early universe, $v$ as a function of the temperature, 
is expected to vanish resulting in a scale-invariant $R^2$ action. 
This model prevents the appearance of a cosmological constant usually 
arising in such cases \cite{Linde}.
The replacement of $\varphi$ by a scalar $\bar\psi\psi$ proposed by Adler 
 does not improve the situation \cite{Adl}.

Here we show that this problem is overcome by the substitution  of the 
tensor field $l_{\mu\nu}$ for the scalar $\varphi$ and the use of the brane 
action $S_{Dir}|_{p=3}$ (\ref{actnl}), comprising potential (\ref{Vred3}), 
instead of action (\ref{sptgrv}). As a result, we find a modified brane potential 
and the corresponding new models of gravity quadratic in curvature. 
These models include the tensor $l_{\mu\nu}$ playing the role of the Brans-Dicke 
scalars  \cite{BD} commonly used in extended models of gravity.
 
These results are based on the following observations. First we note that the fundamental $3$-brane potential (\ref{Vred3}) has the extremal at  $l_{\mu\nu}=0$. This means the absence of  spontaneous breaking of the scaling, because we choose the vev associated 
with $l_{\mu\nu}$ equal to the value of the diff-invariant trace $Spl\equiv l_{\mu\nu}g^{\mu\nu}$  at the extremal. So, the vev turns out to be zero at $l_{\mu\nu}=0$. 
 Secondly, keeping in mind that $Spl$ is a fundamental invariant of embedded hypersurfaces, called mean curvature, we resume that $l_{\mu\nu}=0$ corresponds to a flat h-ws. This  makes us search for a deformation of potential (\ref{Vred3}) to a new one which  has its extremal at $l_{\mu\nu}=l_{o\mu\nu}\neq0$, so that $Spl_{o}=\mu$, 
 where the constant $\mu$ has the dimension $[\mu]=[L^{-1}]$.
 From this we conclude that the required deformation implies the transition from minimal h-ws $\Sigma^{min}_{4}$ to constant mean curvature (cmc) h-ws $\Sigma^{cmc}_{4}$ \footnote{The string model with constant mean curvature worldsheet 
embedded into $\mathbf{R}^{1,2}$ was studied in \cite{BNCh} by the extension of the Nambu-Goto action.}. 
This observation reduces the discussed problem to the construction of a potential $V_{cmc}(l,g)$ which encodes hypersurfaces with the constant mean curvature $Spl=\mu$.
 
The required potential $V_{cmc}(l,g,\mu)$ is constructed in Section 3 extending the discussed mapping to $(p+1)$-dim. hypersurfaces $\Sigma^{cmc}_{p+1}$ with codim 1 and constant mean curvature. 
As a result, we find 
 action $S_{cmc}$ (\ref{actncmsf}) encoding the H-E action with a cosmological constant and equation of motion  (\ref{eqr1m}) self-consistent with the cmc condition.
We note that the presence of the coupling constants $T_{p}$ and $\mu$ in the 
deformed $p$-brane action controls the dynamical regimes produced by the elastic 
and gravitational forces \footnote{A similar dynamics is observed for string 
in curved space \nocite{RZh, RZh'} [53, 54].}.	 

Studied in Section 4 is the Cauchy problem for the EOM (\ref{eqr1m}) 
which are the second order PDEs. We prove that the Peterson-Codazzi (P-C) 
embedding conditions (\ref{ccr}) and the cmc conditions, spontaneously breaking
the scale symmetry for $3$-branes, can be chosen as the initial data constraints 
preserved by the PDEs (\ref{eqr1m}). Then the Cauchy-Kowalevskaya theorem of uniqueness 
ensures that the solution of Eqs. (\ref{eqr1m}) describes the cmc hyper-worldsheet 
encoded by deformed potential $V_{cmc}(l,g,\mu)$  (\ref{solVcmc}). 

In Section 5 potential $V_{cmc}(l,g,\mu)$ is converted to the Lagrangian of $R^2$ gravity (\ref{cub'}) including the H-E term in addition to pure $R^2$. 
Nowadays there is a great interest to such Lagrangians extending the model \cite{Strb} 
to scale-invariant models (see e.g.  \nocite{CoVen, GorT, PST, BaKL, KHPRRSS, Rin, Rin'} [55 - 61]). 
The latter comprise scale-invariant combinations built from $R$, $R^2$, 
an interaction potential of massless scalar field(s), and a few of dimensionless parameters. 
The spontaneous breaking of the scale-symmetry follows from the solutions
 of the equations of motion including a special scalar potential. 
 Interest in such scale-invariant models is due to the fact that they describe inflation and reheating in our 4-dim.
  Universe and fit the new experimental data from Planck \cite{Planck}. 
A remarkable feature of the modern stage of experimental cosmology is the possibility to select  the relevant models to understand the structure of inhomogeneities in the mass distribution in very small time scales. The above mentioned $R^2$ scale-invariant models study scalar perturbations of the gravitational potential. In the case of $p$-branes we introduce the tensor perturbations associated with the massless symmetric tensor field $l_{\mu\nu}$, and have no any free dimensionless parameters. A massless scalar field $\phi$ naturally appears in the brane scenario in the form of the diff invariant $\phi\equiv Sp\,l$ treated as a dynamical
 mean curvature field. Spontaneous breaking of the scaling for $3$-branes is achieved 
by imposing the condition $<\phi>_{0}= <Sp\,l>_{0}=\mu$ which generates a fundamental mass scale similarly to the Higgs effect in QFT. So, the use of $l_{\mu\nu}$ proposes new scale-invariant tensor models of $R^2$ gravity alternative to the well-known models with scalar fields. This motivates application of the considered brane models to analysis of experimental cosmological data. 

In Section 6 we derive the equation for extremals of 
$V_{cmc}(l,g,\mu)$ and find its general 
solution. 
Consistency of the solution with the Gauss and 
P-C embedding conditions results in the 
matrix equation for null vectors of the  $(p+1)$-dim. 
Einstein tensor (matrix) $G_{\mu\nu}$. 
Its solution reveals the presence of the extremal 
$l_{o\mu\nu}(\xi)= \frac{\mu}{p+1}g_{\mu\nu}(\xi)$.  
 This critical point corresponds to the $AdS_{p+1}$ 
 hyper-ws $\Sigma^{o}_{p+1}$ of constant negative curvature.
Next, the behavior of $V_{cmc}$ in the vicinity
of $l_{o\mu\nu}(\xi)$ is analysed. 
The second partial derivatives of $V_{cmc}$ with respect to $l$
at $l_{o}(\xi)$ have different signs depending, in particular, on the brane
dimension $p$.  So, for string the extremum is the minimum 
that shows classical stability of the string potential in $\mathbf{R}^{1,2}$.
But for cmc membrane ($p=2$) the critical point $l_{o}(\xi)$ is proved to be 
 the  saddle 3-dim. hypersurface. It points to the classical instability 
 of the cmc membrane potential in 4-dim. Minkowski space-time. 
 The classical instability of the fundamental (super)membrane \cite{dWLN},
  described by the potential $V(x(\xi))$  \cite{GoHo}, is well known.
Thus, the transition to 
the hyper-worldsheets with constant mean curvature embedded into $\mathbf{R}^{1,3}$ does not remove the instability.

\section{Branes with minimal hyper-worldsheets}

	In the geometrical approach  Dirac $p$-brane embedded into  $\mathbf{R}^{1,D-1}$ with the  
signature $(+-...-)$ 
are described as dynamical systems with the Poincare symmetry $ISO(1, D-1)$
 spontaneously broken to $ISO(1,p)\times SO(D-p-1)$.
The $p$-branes sweep  minimal
h-ws $\Sigma^{min}_{p+1}$ with zero mean curvature 
\begin{eqnarray}\label{minco}
Spl^{a}:= g^{\mu\nu}l^{a}_{\nu\mu}=0,  
\end{eqnarray}
where  $g_{\mu\nu}$ is the induced  metric in $\Sigma^{min}_{p+1}$. 
 The symmetric tensor $l_{\mu\nu}^{a}(\xi)$ is the second fundamental form 
of the h-ws treated as a constrained multiplet 
of the local group $SO(D-p-1)$. We explain  the constraints  (\ref{minco})
as the inverse Higgs effect.
The indexes $a,b=p+1,p+2,..., D-1$ enumerate   
the orts $\mathbf{n}_{a}(\xi^{\rho})$ of an orthonormal moving frame
 attached to $\Sigma^{min}_{p+1}$ and orthogonal to it. 
 The internal coordinates $\xi^{\mu}=(\tau,\sigma^r), \, r=1,2,..,p$ 
 parametrize the h-ws world vector $\mathbf{x}(\xi^{\rho})$. 
The mapping  of the Dirac p-brane action to the diff and $SO(D-p-1)$ invariant 
h-ws action is realized by the field action \cite{Z_rmp} 
 \begin{eqnarray} 
S_{Dir}= \frac{1}{k_{p}^{2}}\int d^{p+1}\xi\sqrt{|g|}\, 
\{- \frac{1}{4}Sp(H_{\mu\nu}H^{\nu\mu}) \nonumber \\
+ \, \frac{1}{2}\nabla_{\mu}^{\perp}l_{\nu\rho a}\nabla^{\perp \{\mu}l^{\nu\}\rho a}
-\nabla_{\mu}^{\perp}l^{\mu}_{\rho a}\nabla_{\nu}^{\perp}l^{\nu\rho a} + V(l, g)\}, 
\label{actnl}
\end{eqnarray}
where the brackets  $\{\mu\nu\},\, [\mu\nu]$ imply the $\mu,\nu$ 
symmetrization and antisymmetrization, respectively.
The potential energy term $V(l, g)$ is equal to
\begin{eqnarray}\label{solVl}
V(l, g)=V_{Dir}:= - \frac{1}{2} Sp(l_{a}l_{b}) Sp(l^{a}l^{b})
+ Sp(l_{a}l_{b}l^{a}l^{b}) - Sp(l_{a}l^{a}l_{b}l^{b})+ {\rm c_{p}}. 
\end{eqnarray}
 The kinetic term for the  $SO(D-p-1)$ gauge 
field $B_{\mu}^{ab}(\xi)=-B_{\mu}^{ba}(\xi)$ 
in the fundamental representation is given by the trace
 $Sp(H_{\mu\nu}H^{\nu\mu}):=H_{\mu\nu}^{ab}H^{\nu\mu}_{ba}$ 
\begin{eqnarray}\label{cH2}
H_{\mu\nu a}{}^{b}:=(\partial_{[\mu}B_{\nu]} + [B_{\mu}, B_{\nu}])_{a}{}^{b}.
\end{eqnarray} 
  The metric and Y-M covariant derivative $\nabla_{\mu}^{\perp}l_{\nu\rho}{}^{a}$ 
is defined by the expression 	
\begin{eqnarray}\label{cdl}
\nabla_{\mu}^{\perp}l_{\nu\rho}{}^{a}:= \partial_{\mu}l_{\nu\rho}{}^{a}
- \Gamma_{\mu\nu}^{\lambda} l_{\lambda\rho}{}^{a} 
-\Gamma_{\mu\rho}^{\lambda} l_{\nu\lambda}{}^{a} + B_{\mu}^{ab}l_{\nu\rho b}.
 \end{eqnarray} 
The hyper-worldsheet metric $g_{\mu\nu}(\xi)$ in (\ref{actnl}) 
is treated as a background field since its equations 
are encoded by the Gauss's Theorema Egregium 
\begin{eqnarray}\label{cRl}
R_{\mu\nu}{}^{\gamma}{}_{\lambda}=l_{[\mu}{}^{\gamma a} l_{\nu]\lambda a}
\end{eqnarray}
 used under construction of action (\ref{actnl}).
 The Gauss eqs. (\ref{cRl}) together with the Peterson-Codazzi 
 equations  for the covariant derivatives (\ref{cdl}) 
\begin{eqnarray}\label{cr}
\nabla^{\perp}_{[\mu}l_{\nu]\rho}^{a}=0 \ \ \ \ \ \
\end{eqnarray}
 form the consistency conditions for embedding of h-ws into the Minkowski space.
 Below we apply the discussed gauge approach to descrption of  
 hyper-ws with constant mean curvature and derive their potential $V_{cmc}$ (\ref{solVcmc}).

\section {Brane h-ws with constant mean curvature}

For hyper-ws of codim 1 action (\ref{actnl}) with zero gauge field $B_{\mu}^{ab}$ 
(since $a=b=p+1$) encodes quadratic curvature gravity on the hyper-ws with
 zero mean curvature. The reduced potential (\ref{solVl}) found in \cite{Z_mpla} is expressed through $R$
\begin{eqnarray}\label{Vred}
V_{Dir}= - \frac{1}{2}Sp(l^{2})Sp(l^{2})+ {\rm c_{p}}=-\frac{1}{2}R^{2} + {\rm c_{p}}, \,\,\, 
R:= g^{\mu\nu}R_{\mu\nu}.
\end{eqnarray}
 In the case $p=3, D=5$ the coupling $k_{3}$ is dimensionless and reduced action (\ref{actnl}) including $V_{Dir}$ (\ref{Vred}) is invariant under the rigid scale transformations 
\begin{eqnarray}\label{scale}
\xi^{\mu\prime}=e^{-\lambda}\xi^{\mu},  \,  \,  \,  g^{\prime}_{\mu \nu}(\xi^{\prime})=g_{\mu\nu}(\xi), \,  \, \, l^{\prime}_{\mu\nu}(\xi^{\prime})= e^{\frac{p+1}{4}\lambda} l_{\mu\nu}(\xi)
\end{eqnarray}
if a h-ws cosmological constant $c_{3}=0$. Minimal h-ws correspond 
to the vev $v\equiv Spl_{0}=0$. 
Here we derive the deformed potential $V_{cmc}$ generating a non-zero $v=\mu$.
This potential encodes constant mean curvature hyper-ws $\Sigma^{cmc}_{p+1}$ of codim 1. 
To find  $V_{cmc}$ we explore the  hyper-ws action 
\begin{eqnarray}\label{actncmc}
S =\frac{1}{k_{p}^{2}}\int d^{p+1}\xi\sqrt{|g|}
(\frac{1}{2}\nabla_{\mu}l_{\nu\rho}\nabla^{\{\mu}l^{\nu\}\rho}
-\nabla_{\mu}l^{\mu}_{\rho}\nabla_{\nu}l^{\nu\rho} -V(l,g)),
  \end{eqnarray}
 where the reduced field $l_{\mu\nu}:=-l_{\mu\nu}^{(p+1)}$,  
  and obtain the following EOM 
\begin{eqnarray}\label{eqrl}
\frac{1}{2}\nabla_{\mu}\nabla^{[\mu}l^{\{\nu]\rho\}}
=-[\nabla^{\mu},  \nabla^{\{\nu}] l_{\mu}{}^{\rho\}}
- \frac{\partial {V}}{\partial l_{\nu\rho}}.   
\end{eqnarray}
Taking into account the Peterson-Codazzi (P-C) conditions (\ref{cr}) with $B_{\mu}=0$  
\begin{eqnarray}\label{ccr}
\nabla_{[\mu}l_{\nu]\rho}=0 \ \ \ \ \ \
\end{eqnarray}
in (\ref{eqrl}),  vanishes their l.h.s. and  
yields the consistency condition for V(l, g)
\begin{eqnarray}\label{eqV} 
 \frac{\partial {V}}{\partial l_{\nu\rho}}
=-[\nabla^{\mu},  \nabla^{\{\nu}] l_{\mu}{}^{\rho\}}.        
\end{eqnarray}
Using the Bianchi identities for the commutator in  (\ref{eqV}) 
\begin{eqnarray}\label{BI}
[\nabla_{\mu} , \, \nabla_{\nu}] l^{\gamma\rho}
=R_{\mu\nu}{}^{\gamma}{}_{\lambda} l^{\lambda\rho}  
+ R_{\mu\nu}{}^{\rho}{}_{\lambda} l^{\gamma\lambda}   
\end{eqnarray}
and the Gauss eqs. (\ref{cRl}) 
we present the r.h.s. of Eqs. (\ref{eqV}) in the form
 \begin{eqnarray}\label{BIsim}
-\frac{1}{2}[\nabla^{\mu} , \, \nabla^{\{\nu}]l_{\mu}^{\rho\}}
=(l^{2})^{\nu\rho} Spl - l^{\nu\rho} Sp(l^2),
\end{eqnarray}
where $Sp(l^{2})=l_{\mu\rho}l^{\rho}_{\nu}g^{\mu\nu}$. 
Substitution of (\ref{BIsim}) into (\ref{eqV}) 
yields the PDE's 
\begin{eqnarray}\label{eqVl}
\frac{1}{2} \frac{\partial {V}}{\partial l_{\nu\rho}}
=(l^{2})^{\nu\rho}Spl -l^{\nu\rho}Sp(l^{2}).
\end{eqnarray}
for $V$ which have the general solution including a cosmological constant $c_{p}$ 
\begin{eqnarray}\label{genrsol}
V(l,g)=\frac{2}{3}Spl Sp(l^{3}) - \frac{1}{2} Sp(l^{2}) Sp(l^{2}) +c_{p}
\end{eqnarray}
as an integration constant. 
Action (\ref{actncmc}) with $V$ (\ref{genrsol}) takes the form 
\begin{eqnarray}\label{actnscale}
S_{df} =\frac{1}{k_{p}^{2}}\int d^{p+1}\xi\sqrt{|g|}
(\frac{1}{2}\nabla_{\mu}l_{\nu\rho}\nabla^{\{\mu}l^{\nu\}\rho}  \\ 
-\nabla_{\mu}l^{\mu}_{\rho}\nabla_{\nu}l^{\nu\rho} -
\frac{2}{3}SplSp(l^{3}) + \frac{1}{2}Sp(l^{2})Sp(l^{2}) - {\rm c_{p}}) 
\nonumber 
 \end{eqnarray}
which is scale-invariant for $p=3$ and $c_{3}=0$. 
Then the scale-symmetry of $S_{df}$, treated as the gravity action,
 explains the observed smallness of the cosmological constant of our $4$-dim. world 
interpreted as a $3$-brane.
 In other words, the exact scaling requires $c_{3}=0$.  
Thereat, a soft breaking of the scale symmetry corresponds to the consideration of $c_{3}$ 
 as a small parameter. 
Action (\ref{actnscale}) generalizes well-known scale-invariant models of gravity extended by scalar fields. The latter are supported by current experimental data. 
Alternatively,  the scale symmetry can be broken by imposing the condition 
\begin{eqnarray}\label{brcale}
Spl\equiv l_{\nu\rho}g^{\nu\rho}=\mu,
\end{eqnarray}
including  a dimensionfull constant  $\mu$. 
As a result, we obtain the potential 
\begin{eqnarray}\label{solVcmc} 
V=V_{cmc}:=\frac{2}{3}\mu Sp(l^{3}) - \frac{1}{2} Sp(l^{2}) Sp(l^{2}) +c_{p} \, , 
 \ \ \ \   Spl=\mu(p)
\end{eqnarray}
which encodes h-ws with the constant mean curvature (cmc) equal to $\mu(p)$.
 Then the action (\ref{actncmc}) is represented in the following form  
\begin{eqnarray}\label{actncmsf}
S_{cmc} =\frac{1}{k_{p}^{2}}\int d^{p+1}\xi\sqrt{|g|}
(\frac{1}{2}\nabla_{\mu}l_{\nu\rho}\nabla^{\{\mu}l^{\nu\}\rho}  \\ 
-\nabla_{\mu}l^{\mu}_{\rho}\nabla_{\nu}l^{\nu\rho} -
\frac{2}{3}\mu Sp(l^{3}) + \frac{1}{2}Sp(l^{2})Sp(l^{2}) - {\rm c_{p}}). 
\nonumber 
 \end{eqnarray}
This action is consistent with the Gauss (\ref{cRl}) and P-C (\ref{ccr}) embedding conditions.  
From the viewpoint of the variational principle, the hyper-worldsheet metric $g_{\mu\nu}$
 works as a background metric, since its evolution is defined by the Gauss eqs.(\ref{cRl}). 
The Euler-Lagrange equations  for $l_{\mu\nu}$ following from (\ref{actncmsf}) are 
\begin{eqnarray}\label{cmceqs}
\frac{1}{2}\nabla_{\mu}\nabla^{[\mu}l^{\{\nu]\rho\}}=2(l^2)^{\nu\rho}(Spl-\mu)
\end{eqnarray}
 and their consistency with (\ref{ccr}) demands the condition $(l^2)^{\nu\rho}(Spl-\mu)=0$ 
 equivalent to  $(Spl-\mu)=0$ \footnote{The second solution $(l^2)^{\nu\rho}=0$
means $Sp(l^3)=Sp(l^2)=0$ that corresponds to $V=c_{p}$.}.
Then Eqs. (\ref{cmceqs}) are reduced to the eqs. 
\begin{eqnarray}\label{eqr1m}
\frac{1}{2}\nabla_{\mu}\nabla^{[\mu}l^{\{\nu]\rho\}}=0, \, \, \, \, \, \, \,   Spl=\mu.
\end{eqnarray}
Using the identity $\nabla^{\perp[\mu}l^{[\nu]\rho]}=-\nabla^{\perp[\nu}l^{\rho]\mu}$ 
transforms  (\ref{eqr1m}) into the equation
\begin{eqnarray}\label{eqgW2sr} 
\frac{1}{2}\nabla_{\mu}\nabla^{[\mu}l^{\{\nu]\rho\}}\equiv
\nabla_{\mu}(\nabla^{[\mu}l^{\nu]\rho} 
 + \frac{1}{2}\nabla^{[\nu}l^{\rho]\mu})=0.
 \end{eqnarray}
Contracting $(\ref{eqgW2sr})$ with $g_{\rho\nu}$ and using the condition $Spl=\mu$ 
 we obtain the covariant conservation law for the divergence  $\nabla^{\rho}l_{\rho}{}^{\mu}$
\begin{eqnarray}\label{div}  
 \nabla_{\mu}(\nabla^{\rho}l_{\rho}{}^{\mu})=0.
\end{eqnarray}
Now it is suitable to rewrite EOM (\ref{eqgW2sr}) in the equivalent form 
$$
\nabla_{\mu}\nabla^{[\mu}l^{\nu]\rho} =-\frac{1}{2}
\{ [\nabla^{\mu}, \nabla^{[\nu}]l^{\rho]}{}_{\mu} +\nabla^{[\nu}\nabla_{\mu}l^{\rho]}{}^{\mu}
\}.
$$
Observing that the $(\nu,\rho)$-antisymmetric 
part of the commutator is presented in the form
\begin{eqnarray}\label{BIasim}
-\frac{1}{2}[\nabla^{\perp\mu} , \, \nabla^{\perp[\nu}]l_{\mu}^{\rho]a}
=\frac{1}{2}([l^{a}, l^{b}])^{\nu\rho} Spl_{b},
\end{eqnarray}
as it follows from  the Bianchi identities and Gauss conditions, 
we obtain vanishing of the l.h.s. of (\ref{BIasim}) for codim 1. 
Then EOM (\ref{eqr1m}) are reduced to 
\begin{eqnarray}\label{eqr1M}
\nabla_{\mu}\nabla^{[\mu}l^{\nu]\rho} + 
\frac{1}{2}\nabla^{[\nu}\nabla_{\mu}l^{\rho]\mu } =0,  \, \, \ \ \ \ \ Spl=\mu.
\end{eqnarray}
 Eqs. (\ref{eqr1M}) are consistent with Eqs. (\ref{ccr}) 
 in view of the identity
\begin{eqnarray}\label{codcon}
\nabla^{\mu}l_{\mu\lambda}=\nabla_{\lambda}Spl\equiv\partial_{\lambda}Spl=0
\end{eqnarray}
following from (\ref{ccr}) after contraction of 
$\mu$ with $\rho$ and using $Spl=\mu$.  

We note that Eqs. (\ref{eqr1M}) may have other solutions that
 do not satisfy the P-C eqs. (\ref{ccr}) and violate the embedding 
conditions for hyper-worldsheets.
 To select the relevant solutions describing brane h-ws,
 we have to add the appropriate initial data conditions (IDC). 
 The latter are chosen to be the first order PDEs (\ref{ccr}) and 
  the condition $\nabla^{\mu}l_{\mu\lambda}=0$ satisfying the covariant conservation
 law (\ref{div}). In the next section we  prove that the IDC
 form a remarkable set of constraints preserved by 
 the second order PDEs  (\ref{eqr1M}).

\section {Cauchy constraints}

The Cauchy problem for  PDEs  (\ref{eqr1M}) is studied here treating
all eqs. (\ref{ccr}) as the initial data constraints fixed 
at the moment $\tau=0$
\begin{eqnarray}\label{cauchy}
 \nabla^{[\mu}l^{\nu]\rho}(0,\sigma^r)=0,   \ \ \ \
 \ \ \ \ \nabla_{\mu}l^{\mu\rho}(0,\sigma^r)=0. 
\end{eqnarray}
Using the results  \cite{Z_rmp} we prove that the constraints (\ref{cauchy}) 
are always fulfilled  during the evolution prescribed by EOM (\ref{eqr1M}).  
Thereat we take into account vanishing of the space covariant derivatives 
 of the IDC (\ref{cauchy}) 
\begin{eqnarray}\label{cauchy'} 
\nabla_{r}\nabla_{\mu}l^{\mu\rho}|_{\tau=0}=0, \ \ \ \
\nabla_{r}\nabla^{[\mu}l^{\nu]\rho }|_{\tau=0}=0,  \ \ \ \ (r=1,2,...,p).
\end{eqnarray}
Consider the power series expansion of  $\nabla^{[\mu}l^{\nu]\rho}$ 
and $\chi^{\rho}\equiv\nabla_{\mu}l^{\mu\rho}$ 
 \begin{eqnarray}
 \chi^{\rho}(\delta\tau,\sigma^r)=\chi^{\rho}|_{\tau=0} + 
\partial_{\tau}\chi^{\rho}|_{\tau=0}\delta\tau + ... =
\nabla_{\tau}\chi^{\rho}|_{\tau=0}\delta\tau + ...,
 \label{divt}
 \\
 \nabla^{[\tau}l^{\nu]\rho}(\delta\tau,\sigma^r )=
 \partial_{\tau} \nabla^{[\tau}l^{\nu]\rho}|_{\tau=0}\delta\tau +...
 =\nabla_{\tau}\nabla^{[\tau}l^{\nu]\rho}|_{\tau=0}\delta\tau +... \, .
 \label{ccrt}
 \end{eqnarray}
The use of Eqs. (\ref{div}) $\nabla_{\tau}\chi^{\tau}(\xi)=-\nabla_{r}\chi^{r}(\xi)$ 
and (\ref{cauchy'}) $\nabla_{r}\chi^{r}|_{\tau=0}=0$ gives   
\begin{eqnarray}\label{div1} 
 \chi^{\tau}(\delta\tau,\sigma^r)=\chi^{\tau}|_{\tau=0}=0.
\end{eqnarray}
 To prove that $\chi^{r}(\delta\tau,\sigma^r)=0$ we consider 
EOM (\ref{eqr1M}) for $\nu=\tau$ and  $\rho=r$ 
\begin{eqnarray}\label{eqr1Ma} 
\nabla_{s}\nabla^{[s}l^{\tau]r} + \frac{1}{2}\nabla^{\tau}\chi^{r}= 0,
 \end{eqnarray}
and taking into account (\ref{cauchy'}), (\ref{div1}) find 
\begin{eqnarray}\label{div2} 
\nabla_{\tau}\chi^{r}|_{\tau=0}=\partial_{\tau}\chi^{r}|_{\tau=0}=0 
\longrightarrow   \chi^{r}(\delta\tau,\sigma^r)=\chi^{r}|_{\tau=0}=0.
\end{eqnarray}
Eqs. $\partial_{\tau} \chi^{\rho}|_{\tau=0}=0$ and (\ref{div1}) 
complete the proof of the desired statement 
\begin{eqnarray}\label{divf} 
\nabla_{\mu}l^{\mu\rho}(\xi)=0.
\end{eqnarray}
Eqs. (\ref{divf}) prove equivalence of EOM (\ref{eqr1M}) to the system 
\begin{eqnarray}\label{eqr1F}
\nabla_{\mu}\nabla^{[\mu}l^{\nu]\rho} =0, \ \ \  \nabla_{\mu}l^{\mu\rho}(\xi)=0
 \end{eqnarray}
showing the local conservation of the P-C conditions (\ref{ccr}). 

Next we prove that the IDC (\ref{cauchy}) together with 
the local conservation laws (\ref{eqr1F}) provide fulfillment of the P-C 
conditions (\ref{ccr}) at any $\tau$
\begin{eqnarray}\label{ccrglb}
\nabla^{[\mu}l^{\nu]\rho}(\xi) =0.
\end{eqnarray}
To this objective consider  $\nu=r$ and $\rho=\tau$ in (\ref{eqr1F}) 
and obtain the relations  
\begin{eqnarray}\label{eqr1Fa}
\nabla_{\tau}\nabla^{[\tau}l^{r]\tau}=- \nabla_{s}\nabla^{[s}l^{r]\tau}
 \end{eqnarray}
which in combination with the IDC (\ref{cauchy'}) show that 
\begin{eqnarray}\label{eqr1Fa'}
\nabla_{\tau}\nabla^{[\tau}l^{r]\tau}|_{\tau=0}=
\partial_{\tau}\nabla^{[\tau}l^{r]\tau}|_{\tau=0}=0.
 \end{eqnarray}
 The substitution of (\ref{eqr1Fa'}) into the expansion (\ref{ccrt}) 
proves that 
\begin{eqnarray}\label{eqr1F1}
\nabla^{[\tau}l^{r]\tau}(\xi) =0.
\end{eqnarray}
The choice of both indices $\nu, \rho $ in (\ref{eqr1F}) as space 
ones $\nu=r, \rho=s$ gives    
\begin{eqnarray}\label{eqr1Fb}
\nabla_{\tau}\nabla^{[\tau}l^{r]s}=- \nabla_{q}\nabla^{[q}l^{r]s}.
 \end{eqnarray}
 Eqs. (\ref{eqr1Fb}) and (\ref{cauchy'}) show that
$\nabla_{\tau}\nabla^{[\tau}l^{r]s}|_{\tau=0}=\partial_{\tau}\nabla^{[\tau}l^{r]s}|_{\tau=0}=0$ that provides 
\begin{eqnarray}\label{eqr1F2}
\nabla^{[\tau}l^{r]s}(\xi) =0. 
\end{eqnarray}
 Eqs. (\ref{eqr1F2}) and (\ref{eqr1F1}) testify to fulfillment of the subset
\begin{eqnarray}\label{eqr1F3}
 \nabla^{[\tau}l^{r]\rho}(\xi) =0
\end{eqnarray}
of the P-C equations.  
 Antisymmetrization in the $\rho, r$ indices in (\ref{eqr1F3}) 
 shows fulfillment of the next subset of Eqs. (\ref{ccr})
\begin{eqnarray}\label{eqr1F3'}
\nabla^{[r}l^{s]\tau}(\xi) =0.
\end{eqnarray}
The remaining subset of the P-C eqs. to be proved
\begin{eqnarray}\label{eqr1Fsp}
\nabla^{[r}l^{s]q}(\xi) =0
\end{eqnarray}
 contains only space-like  indices $r, s, q$. 
 Therefore, we consider the expansion 
\begin{eqnarray}\label{ccrt1} 
 \nabla^{[r}l^{s]q}(\delta\tau,\sigma^r)= \partial_{\tau}\nabla^{[r}l^{s]q}|_{\tau=0}\delta\tau + ...=
g_{\tau\tau}\nabla^{\tau}\nabla^{[r}l^{s]q}|_{\tau=0}\delta\tau + ...,
 \end{eqnarray}
where the relation $\nabla_{\tau}\nabla^{[r}l^{s]q}|_{\tau=0}=g_{\tau\tau}\nabla^{\tau}\nabla^{[r}l^{s]q}|_{\tau=0}$ 
together with  (\ref{cauchy}), (\ref{cauchy'}) are used.
 The proof of  (\ref{eqr1Fsp}) reduces to the that of the relation
\begin{eqnarray}\label{ccrt1c} 
\nabla^{\tau}\nabla^{[r}l^{s]q}|_{\tau=0}\equiv([\nabla^{\tau},\nabla^{[r}]l^{s]q} + \nabla^{[r|}\nabla^{\tau}l^{q|s]})|_{\tau=0}=0.
\end{eqnarray}
Using the identities (\ref{BI}) we present the first term in the r.h.s. of (\ref{ccrt1c}) as  
\begin{eqnarray}\label{ccrt1c1}
[\nabla^{\tau},\nabla^{[r}]l^{s]q}= R^{\tau[rs]}{}_{\mu}l^{\mu q} + R^{\tau[r|q}{}_{\mu}l^{|s]\mu}.
\end{eqnarray}
The conditions (\ref{eqr1F2}) permit to rewrite  the second term in the form 
\begin{eqnarray}\label{ccrt1c2}
\nabla^{[r|}\nabla^{\tau}l^{q|s]}=\nabla^{[r|}\nabla^{q}l^{\tau|s]} = [\nabla^{[r|}, \nabla^{q}]l^{\tau|s]} +\nabla^{q}\nabla^{[r}l^{s]\tau}=\\ 
R^{[r|q\tau}{}_{\mu}l^{\mu|s]} + R^{[r|q|s]}{}_{\mu}l^{\tau\mu} + \nabla^{q}\nabla^{[r}l^{s]\tau}.
\nonumber
\end{eqnarray}
The sum  of (\ref{ccrt1c1}) and (\ref{ccrt1c2}) transformss (\ref{ccrt1c}) into
\begin{eqnarray}\label{ccrt1c'}
\nabla^{\tau}\nabla^{[r}l^{s]q}|_{\tau=0}=(R^{\tau[rs]}{}_{\mu}l^{\mu q} 
+ R^{\tau[r|q|}{}_{\mu}l^{s]\mu} +R^{[r|q\tau}{}_{\mu}l^{\mu|s]} + R^{[r|q|s]}{}_{\mu}l^{\tau\mu})|_{\tau=0}
\end{eqnarray}
after using (\ref{eqr1F3}) and (\ref{cauchy'}).
 Further, we apply the Gauss theorem (\ref{cRl})
\begin{eqnarray}\label{cR1'}
R^{\gamma\lambda\nu}{}_{\mu}l^{\mu r}= (l^{2})^{r[\gamma }l^{\lambda]\nu}
\end{eqnarray}
 and obtain the following expressions for the terms in (\ref{ccrt1c'})
\begin{eqnarray}
R^{\tau[rs]}{}_{\mu}l^{\mu q}=- (l^{2})^{q[r}l^{s]\tau}, \ \ \  R^{[r|q|s]}{}_{\mu}l^{\mu\tau}=(l^{2})^{\tau[r}l^{s]q}, 
\label{G1}  \\
R^{\tau[r|q|}{}_{\mu}l^{s]\mu } + R^{[r|q\tau|}{}_{\mu}l^{s]\mu}  
= (l^{2})^{\tau[s}l^{r]q} -(l^{2})^{q[s}l^{r]\tau}. 
\label{G2}
\end{eqnarray}
Here we observe mutual cancellation of the corresponding terms 
in the sum of (\ref{G1}) and (\ref{G2}). 
This proves the validity of Eqs. (\ref{eqr1Fsp}) and (\ref{ccrglb}).
Then the cmc condition  $Spl=\mu$ emerges as a consequence 
of Eqs. (\ref{divf}) and (\ref{ccrglb}).

So, we show that EOM (\ref{eqr1m}) together with Eqs. (\ref{cRl})
provide conservation of IDC (\ref{divf}) and (\ref{ccrglb}). 
The latter select a {\it closed} sector of the solutions descrbing 
constant mean curvature hyper-ws of codim 1. 

Below, in Section 5, we  prove that the $p$-brane action $S_{cmc}$ (\ref{actncmsf})
 encodes the gravity action  quadratic in curvature and including the desired H-E term.

\section {Hilbert-Einstein action with $R^2$ term}

The Gauss's theorem (\ref{cRl}) permits to express $V_{cmc}$ (\ref{solVcmc}) 
associated with the constant mean curvature hyper-ws $\Sigma^{cmc}_{p+1}$ 
in terms of the Riemann tensor
\begin{eqnarray}\label{cub}
R_{\nu\lambda}=(l^{2})_{\nu\lambda} - \mu l_{\nu\lambda}, \ \
R_{\nu\lambda}l^{\lambda\nu}= Sp(l^{3})-\mu Sp(l^{2}), \ \ R=Sp(l^{2})-\mu^{2}, 
\end{eqnarray}      
where $R_{\nu\rho}$ is the Ricci tensor. 
Using (\ref{cub}) we obtain the  $R$-reps. for $V_{cmc}$  		
\begin{eqnarray}\label{cub'}
V_{cmc}= -\frac{1}{2}R^{2} - \frac{\mu^{2}}{3}R 
+ \frac{2\mu }{3}R_{\nu\lambda}l^{\lambda\nu} + \frac{\mu^{4}}{6} + c_{p}
\end{eqnarray} 
 which reduces to $V_{Dir}$ (\ref{Vred}) when $\mu=0$
\begin{eqnarray}\label{HEcub}
V_{cmc}=V_{Dir}-\frac{\mu^{2}}{3}(R - \frac{\mu^{2}}{2}
- \frac{2}{\mu} R_{\nu\lambda}l^{\lambda\nu}). 
\end{eqnarray} 

Representation (\ref{cub'}) can be rewritten in the equivalent form  as
\begin{eqnarray}\label{cub''}
V_{cmc}= -\frac{\mu^{2}}{3} 
\{ (R -2\Lambda) + \frac{3}{2\mu^{2}}R^{2} -\frac{2}{\mu}R_{\nu\lambda}l^{\nu\lambda}\}
\end{eqnarray}
 after unification of two independent cosmological terms into cosmological
constant $\Lambda(p):=(\frac{\mu^{2}(p)}{4} + \frac{3c_{p}}{2\mu^{2}(p)})$. 
 Due to arbitrariness of $c_{p}$ the constant $\Lambda$ is additional 
dimensionfull phenomenological  parameter of the quadratic gravity derived from $p$-branes.
In particular, one can choose $\Lambda =0$. The potential $V_{cmc} =\frac{2}{3}\mu^{2}\Lambda(p)$ when $R_{\nu\rho}=0$. 
 Then  (\ref{cub}) yields the relations 
 \begin{eqnarray}\label{ricflat}
(l^{n})_{\nu\rho}=\mu^{n-1}l_{\nu\rho}, \, \,  (n=2,3,...) \Longrightarrow l_{\nu\rho}=\mu g_{\nu\rho}, 
\  detl_{\nu}{}^{\alpha}\neq 0 
\end{eqnarray} 
 which show that $Spl=(p+1)\mu$, otherwise $l_{\nu\rho}=0$. 
Solution (\ref{ricflat}) is compatible with the cmc condition $Spl=\mu$ only if $p=0$ that corresponds to a generate case of a worldline swept by point-like $0$-brane.  
 describes.
 
We note that the H-E Lagrangian in  (\ref{HEcub}) is generated simultaneously 
with the cosmological constant dependent on $\mu$ and the new bilinear invariant 
including the Ricchi tensor. 
This deformation of  $V_{Dir}$ is a result of the transition from minimal h-ws 
to those with a constant mean curvature. 
For $p=3, c_{3}=0$, emergence of the H-E term and others in (\ref{HEcub}) breaks the scale symmetry of the four dimensional $R^{2}$ 
gravity encoded by $V_{Dir}|_{p=3}$  (\ref{Vred3}). 
The generated Newton constant turns out to be equal to $G_{N}\approx\frac{1}{\mu^2}$ 
similarly to the case of spontaneously generated gravity in scale-invariant 
 action (\ref{sptgrv}). This shows that $\mu$ is identified with the vev of $Spl$
considered as a dynamical scalar field  $\phi\equiv Spl$. 
 To see this effect we come back to action (\ref{actnscale}) and potential  (\ref{genrsol})
 represented  in the form of the gravity action Lagrangian
\begin{eqnarray}\label{cubfi}
V_{\phi}= -\frac{1}{2}R^{2} - \frac{1}{3}R\phi^{2} 
+ \frac{2}{3}R_{\nu\lambda}l^{\lambda\nu}\phi  + \frac{1}{6}\phi^{4} + c_{p}
\end{eqnarray} 
after  using (\ref{cRl}).
 Comparison of (\ref{cubfi}) with (\ref{cub'}) shows that $V_{\phi}=V_{cmc}|_{\mu=\phi}$.
For $p=3, c_{3}=0$  the $3$-brane action with $V_{\phi}$ (\ref{cubfi}) is scale-invariant. 
Action Lagrangian (\ref{cub''}) with broken scale-symmetry occurs since potential $V$ (\ref{genrsol}) 
and (\ref{cubfi}), respectively, have the critical point $<\phi>_{0}=\mu$, as shown in Section 6.
Potential (\ref{cubfi}) can be compared with that from the model \cite{Rin} 
describing inflation and reheating in the presence of a massless scalar
 field $\varphi$. 
The latter can be identified with the composite scalar $\phi$ in (\ref{cubfi}) 
associated with brane matter.  
So, the massless tensor perturbations  $l_{\mu\nu}$  result in creation of 
 new models of $R^2$ gravity which can be used for analyzing the of current experiments. 

\section {Extremals of the brane potential $V_{cmc}$}

The above mentioned connection of the brane potential $V_{cmc}$ with inflation models 
puts forward the question on its extremals which are defined by solutions of the equation 
\begin{eqnarray}\label{extrmV}
\frac{1}{2}\frac{\partial {V_{cmc}}}{\partial l_{\nu\rho}}
=\mu(l^{2})^{\nu\rho}- l^{\nu\rho}Sp(l^{2})=0.
\end{eqnarray} 
In the dual $R$-representation  this equation takes the form
\begin{eqnarray}\label{Rextr}
Rl_{\nu\rho}(\xi)=\mu R_{\nu\rho}
\end{eqnarray}
after using relations (\ref{cub}). 
The general solution of (\ref{Rextr}) is 
 \begin{eqnarray}\label{solextr}
l_{\nu\rho}(\xi)=\mu\frac{R_{\nu\rho}}{R},   \ \ \   R\neq 0.
\end{eqnarray}
The substitution of this extremal
in the Gauss eqs. (\ref{cRl}) yields the relation 
\begin{eqnarray}\label{extrGau}
R_{\mu\nu\gamma\lambda}=-(\frac{\mu}{R})^{2}(R_{\mu\gamma} R_{\nu\lambda}-R_{\nu\gamma} R_{\mu\lambda}).
\end{eqnarray} 
The extremal (\ref{solextr}) should obey the Peterson-Codazzi eqs. (\ref{ccr}) 
\begin{eqnarray}\label{ccr*}
\nabla_{[\mu}R_{\nu]\rho}-\frac{1}{R} R_{\rho[\nu} \nabla_{\mu]}R=0.
\end{eqnarray}
and  Eq. (\ref{codcon}) which follows 
from (\ref{ccr}) and has the form   
\begin{eqnarray}\label{ccrcon}
\nabla_{\nu}R^{\nu\rho} - \frac{1}{R}R^{\rho\nu}\partial_{\nu}R =0
\end{eqnarray}
after substitution of (\ref{solextr}) in (\ref{codcon}).
The use of the relation 
\begin{eqnarray}\label{conbia}
\nabla_{\nu}R^{\nu}{}_{\rho} = \frac{1}{2}\partial_{\rho}R,
\end{eqnarray}
which follows from the Bianchi identities 
\begin{eqnarray}\label{bia}
 \nabla_{\alpha} R^{\mu}{}_{\nu\gamma\lambda} + \nabla_{\gamma} R^{\mu}{}_{\nu\lambda\alpha}
  + \nabla_{\lambda} R^{\mu}{}_{\nu\alpha\gamma} =0,
\end{eqnarray}
transforms Eq. (\ref{ccrcon}) into the equation  
\begin{eqnarray}\label{hein}
(R^{\mu\nu} - \frac{1}{2}g^{\mu\nu}R)\partial_{\nu}R=0.
\end{eqnarray}
The latter means that  $\partial_{\mu}R$
should be a null vector for the Einstein 
matrix $G^{\mu\nu}:=R^{\mu\nu} - \frac{1}{2}g^{\mu\nu}R$ 
associated with $\Sigma^{extr}_{p+1}$.  
 In the $R$-reps. the extremal  h-ws is fixed by 
 Eqs. (\ref{extrGau}), (\ref{hein}). Their solution 
  for the case when $detG^{\mu\nu}\neq 0$ is 
 \begin{eqnarray}\label{heincc}
 \partial_{\nu}R=0   \Longrightarrow  R=R_{o}=constant.
 \end{eqnarray}
 This solution defines a h-ws $\Sigma^{o}_{p+1}$ of {\it constant curvature} (cc) 
 characterized by the Riemann tensor and the Gauss curvature $K_{o}$   
 \begin{eqnarray}\label{extrRimo}
R_{o\mu\nu\gamma\lambda}=K_{o}(g_{\mu\gamma} g_{\nu\lambda}-g_{\nu\gamma} g_{\mu\lambda}). 
\end{eqnarray} 
The substitution of  (\ref{extrRimo}) in the Gauss eqs. (\ref{extrGau})             
 results in the relation 
\begin{eqnarray}\label{extrGauK}
K_{o}(g_{\mu\gamma} g_{\nu\lambda}-g_{\nu\gamma} g_{\mu\lambda})
=-(\frac{\mu}{R})^{2}(R_{\mu\gamma} R_{\nu\lambda}-R_{\nu\gamma} R_{\mu\lambda})
\end{eqnarray} 
which shows proportionality of  $R_{o\mu\nu}$ to $g_{\mu\nu}$ and defines $K_{o}$ 
\begin{eqnarray}\label{extrcc}
R_{o\mu\nu}=\Lambda_{o} g_{\mu\nu}, \ \ R_{o}=(p+1)\Lambda_{o}, 
 \ \  K_{o}=-(\frac{\mu}{p+1})^{2}.
\end{eqnarray}  
The latter completely fixes the Riemann tensor of $\Sigma^{o}_{p+1}$ 
\begin{eqnarray}\label{extrRimf}
R_{o\mu\nu\gamma\lambda}
=-(\frac{\mu}{p+1})^{2}(g_{\mu\gamma} g_{\nu\lambda}-g_{\nu\gamma} g_{\mu\lambda}) 
\end{eqnarray} 
and the function $\Lambda_{o}=pK_{o}$, respectively. Then we obtain 
\begin{eqnarray}\label{extrccf}
R_{o\mu\nu}=-\frac{p}{(p+1)^{2}}\mu^{2} g_{\mu\nu}, \ \ \ \ \ 
 R_{o}:=g^{\mu\nu}R_{o\mu\nu}=-\frac{p}{p+1}\mu^{2}.
\end{eqnarray}  
 As shows the substitution of (\ref{extrccf}) in (\ref{solextr}),  
 this extremal acquires the form 
\begin{eqnarray}\label{solextrf}
l_{o\mu\nu}= \frac{\mu}{p+1}g_{\mu\nu} \,  \Longrightarrow \,
 R_{o\mu\nu}=-\frac{\mu p}{p+1} l_{o\mu\nu}.
\end{eqnarray}
Solutions (\ref{solextrf}) are proportional to $g_{\mu\nu}(\xi)$, 
and therefore they are solutions of the P-C eqs. (\ref{ccr}) 
in the $l$-reps.  and Eqs. (\ref{ccr*}) in the $R$-reps., respectively. 
Note that the extremal of $V(l, g)$ (\ref{genrsol})
is $l_{o\mu\nu}= \frac{Sp\,l_{o}}{p+1}g_{\mu\nu}$ which 
satisfies Eqs. (\ref{ccr})
 for $Sp\,l_{o}=\mu$.  

Coming back to Eq. (\ref{extrmV}) in the $l$-reps. we rewrite it in the form 
$$l^{\nu}_{\alpha}(\mu l^{\alpha\rho} - g^{\alpha\rho} Sp(l^{2}))=0$$ 
which shows that there exist two sets of the solutions. 
For the first of them  $detl_{\mu\nu}=0$, for
the other it is nonzero.  
In the latter case (\ref{extrmV}) reduces to 
\begin{eqnarray}\label{extrmV'}
l^{\alpha\rho} =\frac{Sp(l^{2})}{\mu} g^{\alpha\rho},  \ \ \ \  detl_{\mu\nu}\neq 0
\end{eqnarray} 
and corresponds to the above-considered solution (\ref{solextrf}).
Indeed, in this case contraction of (\ref{extrmV'}) with $g^{\rho\alpha}$ 
and $l^{\rho\alpha}$yields the conditions
\begin{eqnarray}\label{extrmrel}
Spl_{o}=\mu, \ \ \ \   Sp(l_{o}^{2})=\frac{\mu}{p+1}Spl_{o} =\frac{\mu^{2}}{p+1},
\end{eqnarray} 
 respectively. 
Then substitution of (\ref{extrmrel}) in (\ref{extrmV'}) reproduces 
solution (\ref{solextrf}) for  $l_{o\mu\nu}$.
Substitution of $l_{o\mu\nu}$ in the Gauss eqs.  (\ref{cRl}) 
 yields solution  (\ref{extrRimf}) for the Riemann tensor 
  of the extremal h-ws $\Sigma^{o}_{p+1}$. 
The h-ws  $\Sigma^{o}_{p+1}$  is a hyperbolic one
with the negative {\it constant curvature} $K_{o}$. 
 The use of the relation 
$$
R_{o\mu\nu}=\frac{R_{o}}{p+1}g_{\mu\nu},  \ \ \ 
\Lambda_{o}=-\frac{p}{(p+1)^{2}}\mu^{2}
$$
shows that $\Sigma^{o}_{p+1}$ is a $(p+1)$-dim. Einstein space
 with the negative cosmological constant $\Lambda_{o}$ which 
can be treated as the anti-de Sitter space $AdS_{p+1}$.

The vacuum value of $V_{cmc}$ (\ref{solVcmc}) is the constant given by 
\begin{eqnarray}\label{extrV}
V_{o}:=V_{cmc}(l=l_{o}, g, \mu)=\frac{\mu^{4}(p)}{6(p+1)^{2}} - c_{p}.
\end{eqnarray}
The action (\ref{actncmsf}) at the extremum $l_{o}(\xi)$ 
is also constant equal to
\begin{eqnarray}\label{actncmsfe}
S_{o}:=S_{cmc}|_{vac} =\frac{1}{k_{p}^{2}}\int d^{p+1}\xi\sqrt{|g|}(\frac{\mu^{4}(p)}{6(p+1)^{2}} - c_{p}),
\end{eqnarray}
 due to vanishing of the kinetic term of $l_{o}(\xi)$ because $l_{o\mu\nu}\sim g_{\mu\nu}$.

The differential of Eq. (\ref{eqVl}) yields the second partial derivative of $V_{cmc}$ 
 \begin{eqnarray}
 \frac{\partial^{2}V_{cmc}}{\partial l_{\alpha\beta}\partial l_{\nu\rho}}
=
[g^{\beta\{\nu}l^{\rho\}\alpha} +  g^{\alpha\{\nu}l^{\rho\}\beta}]Spl - g^{\alpha\{\nu}g^{\rho\}\beta}Sp(l^{2})
\label{deqVl} \\
+ \, \, [(l^{2})^{\nu\rho}g^{\alpha\beta} + (l^{2})^{\alpha\beta}g^{\nu\rho}] - 4l^{\nu\rho}l^{\alpha\beta}. \ \ \ \ \ 
\nonumber 
\end{eqnarray}
The substition of $l_{\mu\nu}=l_{o \mu\nu}$ (\ref{solextrf}) in Eq. (\ref{deqVl}) 
gives its value 
\begin{eqnarray}\label{2deqVl} 
 \frac{\partial^{2}V_{cmc}}{\partial l_{\alpha\beta}\partial l_{\nu\rho}}|_{vac}
=(\frac{\mu}{p+1})^{2}[(p+1)g^{\beta\{\nu}g^{\rho\}\alpha} - 2 g^{\nu\rho}g^{\alpha\beta}]
\end{eqnarray} 
at the critical point $l_{o \mu\nu}(\xi)$. This value  
 is not positively defined that hints on possible classical 
instability of the vacuum. The latter is easily checked 
for strings  and mebranes where the diagonal gauge for $g_{\mu\nu}$ 
can be chosen.  For string ws we obtain two non-zero components 
 $l_{\tau\tau}\equiv l_{\tau}, \, l_{\sigma\sigma}\equiv l_{\sigma}$ that gives
\begin{eqnarray}\label{strin}
\frac{\partial^{2}V_{cmc}}{\partial l_{\tau}\partial l_{\tau}}|_{vac}
=\frac{(\mu g^{\tau})^{2}}{2},
\ \
\frac{\partial^{2}V_{cmc}}{\partial l_{\sigma}\partial l_{\sigma}}|_{vac}
=\frac{(\mu g^{\sigma})^{2}}{2},
\ \ 
\frac{\partial^{2}V_{cmc}}{\partial l_{\tau}\partial l_{\sigma}}|_{vac}
=-(\frac{\mu}{2})^{2} g^{\tau}g^{\sigma}.
\end{eqnarray}
Since $g^{\tau}g^{\sigma}<0$, we find the extremum to be the  
{\it minimum} defining classically stable {\it string} vacuum. 
For  membrane $g_{\tau r}=g_{\sigma\eta}=0$, where 
 $\sigma^{1}\equiv\sigma$ and $\sigma^{2}\equiv\eta$. Then we obtain 
three non-zero diagonal components: 
$l_{\tau\tau}\equiv l_{\tau}, l_{\sigma\sigma}\equiv l_{\sigma}, 
 l_{\eta\eta}\equiv l_{\eta}$,  \
 and Eqs. (\ref{strin}) take the form
 \begin{eqnarray}\label{membrdi} 
\frac{\partial^{2}V_{cmc}}{\partial l_{\tau}\partial l_{\tau}}|_{vac}
=\frac{(2\mu g^{\tau})^{2}}{2},
\ \
\frac{\partial^{2}V_{cmc}}{\partial l_{\sigma}\partial l_{\sigma}}|_{vac}
=\frac{(2\mu g^{\sigma})^{2}}{2},
\ \ 
\frac{\partial^{2}V_{cmc}}{\partial l_{\eta}\partial l_{\eta}}|_{vac}
=\frac{(2\mu g^{\eta})^{2}}{2} 
\end{eqnarray}
with all {\it positive} derivatives. The same we see for the mixed derivatives   
\begin{eqnarray}\label{membrof+}
\frac{\partial^{2}V_{cmc}}{\partial l_{\tau}\partial l_{\sigma}}|_{vac}
=-2(\frac{\mu}{3})^{2} g^{\tau}g^{\sigma},
\ \
\frac{\partial^{2}V_{cmc}}{\partial l_{\tau}\partial l_{\eta}}|_{vac}
=-2(\frac{\mu}{3})^{2} g^{\tau}g^{\eta},
\end{eqnarray}
since $g^{\tau}g^{\sigma}<0$ and $g^{\tau}g^{\eta}<0$ due to the time-like 
character of  $\Sigma^{o}_{3}$.

However, the mixed derivative with respect to $l_{\sigma}$ and $l_{\eta}$ 
is {\it negative}
\begin{eqnarray}\label{membrof-}
\frac{\partial^{2}V_{cmc}}{\partial l_{\sigma}\partial l_{\eta}}|_{vac}
=-2(\frac{\mu}{3})^{2} g^{\sigma}g^{\eta},
\end{eqnarray}
 since $g^{\sigma}g^{\eta}>0$. 
It means that the extremal of the potential $V_{cmc}(l,g,\mu)$ 
is a 3-dim. saddle hypersurface that shows classical {\it instability} of the membrane vacuum.  So, the potential of cmc membrane with $R_{\mu\nu}\neq 0$ in 4-dim. Minkowski 
 space turns out to be instable. This statement generalizes the famous theorem \cite{dWLN} on the classical instability of fundamental (super)membranes to the case of membranes with constant mean curvature of their 3-dim. worldsheets. Our result based on the sign indefiniteness of the second derivatives in (\ref{2deqVl}) hints that the extremals of $V_{cmc}(l,g, \mu)$ for $p$-branes with codim 1 and higher $p>2$ may also be saddle hypersurfaces.

\section{Conclusion}

We studied $p$-branes sweeping hyper-worldsheets (h-ws) of codimension 1 with 
a constant mean curvature $\mu$, and constructed their action expressed through 
the h-ws metric $g_{\mu\nu}$ and Nambu-Goldstone tensor field $l_{\mu\nu}$ 
associated with the broken Poincare symmetry. 
 The interaction potential for  $l_{\mu\nu}$ and $g_{\mu\nu}$ was shown to encode a generalized
 h-ws action of quadratic curvature gravity with a cosmological constant, where $l_{\mu\nu}$ turned out to play the role of a scalar field $\varphi$ similar to the Brans-Dicke one. The latter fulfills  a key function in the Adler-Zee mechanism of spontaneously generated gravity arising from breaking of the scale symmetry \cite{Zee}. We reopened this scenario and its generalization realized by branes, where $<Spl>_{0}\equiv <l_{\mu\nu}g^{\mu\nu}>_{0} =\mu$ replaces the vacuum expectation value 
 $<\varphi>_{0}$. This result follows from the equation for the brane potential extremals and its solution describing h-ws of negative constant curvature. These h-ws are maximally symmetric spaces identified with the anti-de Sitter spaces $AdS_{p+1}$. The found extremal for the membrane potential was proved to be realized by a saddle 3-dim. hypersurface. This points to the classical instability of the membranes with constant mean curvature h-ws in $\mathbf{R}^{1,3}$. This result generalizes the well-known theorem on the classical  instability of fundamental (super)membranes \cite{dWLN}. It is interesting to connect the saddle-like instability of the deformed membranes with the inflation and reheating mechanisms studied in quadratic curvature models of gravity. The obtained results propose new models of $R^2$ gravity with spontaneously broken scale symmetry implementing the symmetric tensor field $l_{\mu\nu}$ instead of well-studied scalar fields.

\vskip 20 pt

\noindent{\bf Acknowledgments}
\vskip 10 pt

The author would like to thank NORDITA and Physics Department 
of Stockholm University for kind hospitality and support, I. Bengtsson,
 S. Lukyanov, H. von Zur-M{\"u}hlen for stimulating discussions, 
G. Huicken for valuable remarks, and A. Zee for his interest to
 the obtained results.

\end{document}